\documentclass[10pt,conference]{IEEEtran}
\usepackage{listings}
\usepackage{xcolor}
\usepackage{graphicx}
\usepackage{caption}
\usepackage{amsmath}
\usepackage{url}

\lstdefinelanguage{LLVM}{
  morekeywords={
    define, declare, global, constant, 
    private, internal, external, 
    linkonce, weak, appending, 
    dllimport, dllexport, common, 
    default, hidden, protected, 
    unnamed_addr, 
    zeroext, signext, inreg, sret, 
    nounwind, noreturn, 
    noalias, nocapture, byval, nest, 
    readnone, readonly, inlinehint, 
    noinline, alwaysinline, optsize, 
    ssp, sspreq, noredzone, noimplicitfloat, naked, 
    module, asm, target, datalayout, triple, 
    blockaddress, 
    type, opaque, 
    ret, br, switch, invoke, unwind, unreachable, 
    add, fadd, sub, fsub, mul, fmul, udiv, sdiv, fdiv, 
    urem, srem, frem, 
    shl, lshr, ashr, and, or, xor, 
    extractelement, insertelement, shufflevector, 
    getelementptr, 
    trunc, zext, sext, fptrunc, fpext, fptoui, 
    fptosi, uitofp, sitofp, ptrtoint, inttoptr, 
    bitcast, addrspacecast, 
    icmp, fcmp, 
    phi, select, call, va_arg, 
    landingpad, 
    catchpad, cleanuppad, catchret, cleanupret, catchswitch,
    catchendpad, 
    alloca, load, store, fence, cmpxchg, atomicrmw, 
    getelementptr, 
    inbounds, to, tail, 
    label, 
    dbgs, dbg, dbginfo, 
    basic_block, 
    function, 
    undefined, null, undef, none, true, false, sam, cmps, nop
  },
  sensitive=false, % LLVM IR is case-insensitive
  morecomment=[l];, % Line comments start with ;
  morestring=[b]", % Strings are enclosed in double quotes
  alsoletter={\%}, % % is considered as a letter for registers
  columns=fullflexible,
  keepspaces=true,
  showstringspaces=false,
  basicstyle=\ttfamily\scriptsize, % Code font style
  keywordstyle=\bfseries\color{blue}, % Keywords style
  commentstyle=\itshape\color{gray}, % Comments style
  stringstyle=\color{orange}, % Strings style
  literate={\%}{\%}1, % Ensure % is not escaped,
}

\begin{document}

\title{Keep Me Updated: An Empirical Study of Proprietary Vendor Blobs in Android Firmware}
\author{
    \IEEEauthorblockN{Elliott Wen}
    \IEEEauthorblockA{The University of Auckland, New Zealand\\
    elliott.wen@auckland.ac.nz}
    \and
    \IEEEauthorblockN{Jiaxing Shen}
    \IEEEauthorblockA{Lingnan University, Hong Kong\\
    jiaxingshen@LN.edu.hk}
    \and
    \IEEEauthorblockN{Burkhard Wuensche}
    \IEEEauthorblockA{The University of Auckland, New Zealand\\
    burkhard@cs.auckland.ac.nz}
}

\maketitle
\begin{abstract}
Despite extensive security research on various Android components, such as kernel or runtime, little attention has been paid to the proprietary vendor blobs within Android firmware.
In this paper, we conduct a large-scale empirical study to understand the update patterns and assess the security implications of vendor blobs. We specifically focus on GPU blobs because they are loaded into every process for displaying graphics user interfaces and can affect the entire system's security. We examine over 13,000 Android firmware releases between January 2018 and April 2024. Our results reveal that device manufacturers often neglect vendor blob updates. About 82\% of firmware releases contain outdated GPU blobs (up to 1,281 days). A significant number of blobs also rely on obsolete LLVM core libraries released more than 15 years ago. To analyze their security implications, we develop a performant fuzzer that requires no physical access to mobile devices. We discover 289 security and behavioral bugs within the blobs. We also present a case study demonstrating how these vulnerabilities can be exploited via WebGL. This work underscores the critical security concerns associated with vulnerable vendor blobs and emphasizes the urgent need for timely updates from device manufacturers.
\end{abstract}
\section{Introduction}
Nowadays, Android plays a fundamental role in our digital lives. It holds a 85\% of the mobile phone market share and is also extensively used in various smart devices, such as TVs and watches. An Android device tends to have a highly complex software composition. Its base is the Linux Kernel, while the system runtime comes from Google's Android Open Source Project (AOSP). Additionally, device manufacturers often pre-install specific system applications and incorporate proprietary binary drivers. This diverse composition results in a fragmented Android ecosystem and poses huge security challenges. 
% This
% success is hugely attributed to the openness of Android, which allows vendors to build their own customized firmware images for different devices based on Android Open Source Project (AOSP). For instance, vendors can integrate their own pre-installed Android Apps to provide a more intuitive user interface. Vendors can customize the firmware by integrating their own pre-installed Android Apps.

% Modern smartphone systems have highly complex software supply chains. For instance, Android, which holds approximately 85\% of the smartphone market, is built on many components. 

% it incorporates proprietary driver blobs from various hardware vendors, such as Qualcomm's GPUs and Sony's cameras.

% We need to talk about more on the vendor driver.

\begin{figure*}[h]
 % \vspace{-15pt}
\centering
\includegraphics[width=0.90\linewidth]{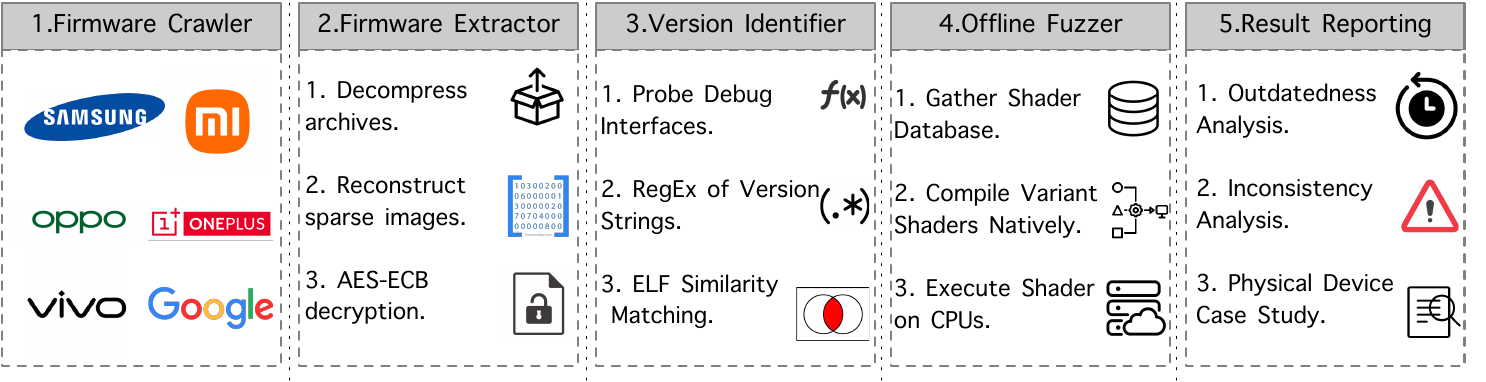}

  \caption{\textit{GPU-Inspector} Pipeline. }
  \label{fig:method}
\vspace{-15pt}

\end{figure*}

Recently, numerous research efforts have been made to investigate  Android component security.  For instance, Zhang et al.~\cite{zhang2021investigation} explored the Android Linux kernel code review process and identified bottlenecks in patch application. Tung et al.~\cite{tung2018android} investigated the device manufacturers' update practices for their AOSP components.
Elsabagh et al.~\cite{elsabagh2020firmscope} applied static analysis to detect privilege escalation vulnerabilities in pre-installed apps. Despite these advancements, a gap remains in the analysis of security situations regarding proprietary vendor blobs within Android firmware. 
These blobs are distributed by device vendors and located in a special firmware partition (i.e., \verb|/vendor|). They provide essential support for the device's hardware components, such as GPUs, cameras, and fingerprint readers. 
Many of them are loaded into apps or system services with high privileges. If these blobs are vulnerable, they can significantly impact the security of the entire system. 
However, due to their closed-source nature and lack of documentation, they are subject to limited scrutiny. 
% a gap remains in the analysis of proprietary driver blobs based on large-scale firmware datasets, and an understanding of the current security situations is still lacking. This may be primarily attributed to the closed-source nature of the proprietary drivers.
 
In this study, we conduct a large-scale empirical evaluation of vendor blobs within Android firmware. Our research aims to 1) understand the update pattern of vendor blobs and 2) identify their vulnerabilities and assess their security implications for the Android ecosystem. We place a specific focus on GPU vendor blobs based on two key reasons. 
Firstly, GPU blobs are more critical to system security than other blobs because every app loads them into memory for rendering graphical user interfaces (UI). 
Secondly, they are easily exploitable; many commonly used apps (e.g., Google Chrome and Firefox) or built-in UI components (e.g., WebKit) can accept arbitrary GPU task inputs and forward them to the vulnerable GPU blobs. 

To achieve our goals, we construct an automatic analysis pipeline named \textit{GPUBlob-Inspector} as shown in Fig~\ref{fig:method}. The pipeline begins with a firmware crawler that gathers Android firmware images from various device manufacturers. The firmware is then input into an image unpacker to extract the proprietary vendor blobs. Afterward, the pipeline proceeds to identify the blob's version number via Executable and Linkable Format (ELF) fingerprints. Subsequently, our pipeline employs a specially designed fuzzer on the GPU vendor blobs to identify potential security vulnerabilities.
Our fuzzer utilizes metamorphic testing; it generates and feeds a series of semantically equivalent GPU programs to the GPU blobs to observe any irregular behaviors or inconsistent program outputs. 
Unlike existing GPU fuzzers that necessitate physical access to mobile devices, our system operates in an offline manner. We achieve this by utilizing the fact that those GPU blobs internally generate Low Level Virtual Machine (LLVM) Intermediate Representation (IR) for the input GPU program. We can capture the LLVM IR and use it to directly generate an executable for performant CPU-based fuzzing.
In the final stage, we conduct a manual analysis to validate the vulnerabilities on physical devices. 

For our study, we curate a large-scale firmware dataset, comprising 13,901 Android images from 74 distinct phone vendors. The data
span a period ranging from Jan 2017 to April 2024. From this dataset, we unveil the following key research findings. 
\begin{enumerate}
\item Most manufacturers neglect timely updates to the GPU vendor blobs; approximately 82\% of firmware
images in our dataset contain outdated GPU blobs. On average, a GPU blob remains outdated for 273 days, with the longest period stretching up to 1,128 days. In addition, certain manufacturers tend to update these blobs far less frequently than others.
\item Most GPU vendor blobs are built upon outdated LLVM core libraries. A significant portion of them rely on \textit{LLVM 2.8}, a version that is released over 15 years ago. Even the most recent LLVM dependency found among these blobs dates back to 5 years ago. 
\item Our offline fuzzer identifies 289 security vulnerabilities and behavioral anomalies within the most recent Qualcomm GPU blobs. These issues stem from memory access violations, endless iterations of optimization passes, and incorrect program semantics. We present a case study to exploit them via WebGL, which may lead to denial of service or arbitrary code execution on a target device.
% We identify and track the version of each GPU blob. 
% We discover that approximately 85\% of mobile devices never update their GPU drivers, meaning these devices continue to use the same outdated engines from their initial release. 
% The driver update frequency is notably less than that of AOSP. 
%  We also observe that certain manufacturers update their proprietary GPU drivers significantly less frequently than others, despite their devices being equipped with identical GPUs.
 % Even within the same manufacturer, different phone models using the same GPU may receive driver updates at different rates. 

% We curate a reference shader database from top-ranking mobile games. These reference shaders are used to generate variant shaders for our metamorphic testing fuzzer. 
% Through this process, we identify a number of security and behavior bugs within the Qualcomm GPU driver, even in the latest versions. Common issues include segmentation faults, incorrect IR code generation, and wrong control flow graph optimizations. A contributing factor is that these drivers internally rely on an outdated version of the LLVM compiler. 
% To highlight the security risks associated with driver vulnerabilities, we carry out a case study on the Chrome browser using WebGL APIs. 
% We showcase two exploitation strategies. 
% The first strategy has the potential to trigger a crash in the browser. The second strategy is capable of accurately monitoring each user's browsing activity, regardless of whether the user has cleared all browser history or is in incognito mode.
\end{enumerate}
To contribute to future research, we also make the source code of our analysis pipeline and the dataset publicly available. 

% Our contributions can be summarized as follows. 
% \begin{enumerate}
% \item Offline Driver Fuzzing Pipeline: We purposed a fuzzing framework, named GFuzzer, to detect potential bugs within proprietary GPU driver binaries. This is accomplished without the need for access to physical devices.
% \item Measurements and Explorations: Based on our extensive data analysis and large-scale fuzzing, we identified numerous driver bugs including segmentation faults, incorrect loop simplifications, and control flow simplifications. Upon discovery, we take the responsible step to notify the relevant vendors about these security issues.
% \end{enumerate}

\section{Background: Proprietary Vendor Blobs}
\begin{figure}[t]
 % \vspace{-15pt}
  \begin{center}
  \includegraphics[width=1\linewidth]{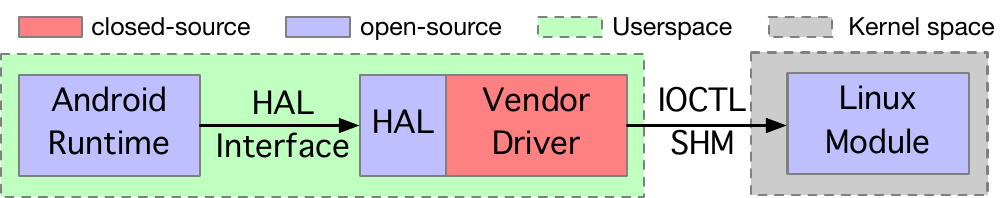}
 % \vspace{-10pt}
  \caption{Proprietary Vendor Blobs in Android.}
  \label{fig:hal}
  \end{center}
 \vspace{-15pt}
  \end{figure}

Android is an open ecosystem, yet it still allows manufacturers to include closed-source blobs in their distributions. 
This practice is typically used for hardware-specific code, where manufacturers may opt to keep their implementation details confidential. These vendor blobs must adhere to a standardized set of interface definitions known as the Hardware Abstraction Layer (HAL), as illustrated in Figure~\ref{fig:hal}. 
% This allows the Android runtime to interact with the blobs without needing to know their specific implementations. 
For instance, a fingerprint reader blob needs to implement the \verb|biometrics.fingerprint@2.1| interfaces, while a GPU blob needs to implement the core OpenGL ES or Vulkan graphics interfaces. There are two types of HAL interfaces: \textit{Binderized} and \textit{Same-process}. A binderized blob (e.g., fingerprint) is generally loaded into a high-privilege system service, while a same-process blob (e.g., GPU) opens in the same process in which it is used. 

Vendor blobs generally operate within the user space and have corresponding Linux modules that run in the kernel space. These blobs and kernel modules communicate through standard input-output subsystem protocols, such as \textit{IOCTL} and shared memory. To comply with the GNU General Public License (GPL) of Linux, the kernel modules are open-source. They generally just perform basic low-level tasks such as register manipulation and memory management, while the core logic for major hardware operations remains implemented and concealed within the vendor blobs.

In the context of GPU blobs, there are two main hardware vendors: Qualcomm Adreno and ARM Mali. They both follow the aforementioned HAL architecture. For example, Qualcomm employs a Linux kernel module named \textit{KGSL} to handle basic low-level hardware operations such as power management and register initialization. Its user-space library \textit{QGL} handles the core GPU tasks, such as providing standard graphics APIs (e.g., OpenGL ES and Vulkan) and compiling GPU programs into hardware instructions. The user-space library is closed-source for device end-users, while its source code is provided to device manufacturers such as Google and Xiaomi under a Non-Disclosure Agreement (NDA). Consequently, the responsibility for updating the driver rests entirely with the manufacturers. 

\section{GPU-Inspector Pipeline}
In this section, we provide implementation details for our \textit{GPUBlob-Inspector} pipeline. 

\subsection{Firmware Curation and Vendor Blob Unpacking}
We implement a web crawler to curate Android firmware images from various vendors. Some vendors, such as Google and Xiaomi, make all historical firmware versions available on their official websites. However, others like Samsung and OnePlus do not provide publicly accessible URLs. In such cases, we source their firmware images from third-party subscription services like SamMobile\footnote{https://www.sammobile.com/} and Daxiaamu\footnote{https://yun.daxiaamu.com
}. Alongside firmware binaries, we also collect metadata information, including firmware version, release date, and supported regions.
To enhance dataset diversity, we also incorporate data from Android Dump, a public repository hosting firmware from lesser-known phone vendors. Our dataset comprises 13,901 firmware images across 24 phone vendors, consuming a total of 38 TBs of disk space. They are released between between January 2018 and April 2024. Among them, the top five phone vendors (Samsung, Xiaomi, Oppo, OnePlus, and Google) collectively contribute to 93\% of the images.

Generally, a firmware image consists of multiple compressed partitions such as \textit{boot}, \textit{recovery}, \textit{system}, and \textit{vendor}. Our goal is to decompress the vendor partition, where the vendor blobs generally reside. To accomplish this, we adapt an Android firmware analysis tool\footnote{https://github.com/srlabs/extractor} as recommended by previous research. Upon decompression, 
we proceed to retrieve the GPU vendor blobs. According to the Android framework specifications, they must be situated within the directory \verb|/vendor/lib64/egl| and may be distributed as a singular binary (\textit{libGLES.so}) or as three separate binaries (\textit{libEGL.so}, \textit{libGLESv1\_CM.so}, and \textit{libGLESv2.so}). 
It is important to note that these GPU libraries may exhibit dependencies on other libraries within the vendor partition, which in turn may have further dependencies. To ensure a thorough extraction of all dependent libraries, we utilize a tool named \textit{readelf}\footnote{https://man7.org/linux/man-pages/man1/readelf.1.html} to parse a library's \textit{DT\_NEEDED} header section and identify its dependencies. We initiate this process with the GPU library and we recursively apply readelf to each dependent library we encounter.
% Based on our observations, ARM typically distributes a single binary, whereas Qualcomm tends to  provide three separate binaries.

\subsection{Driver Version Identification}\label{versionid}
The next task is to identify the version of these GPU vendor blobs. This process is not straightforward due to the absence of version metadata within the binary. Consequently, we devise our own versioning scheme, which comprises the following three components:

\begin{enumerate}
\item Build ID: One rudimentary yet effective versioning approach is to use the build ID of each blob file. A build ID is essentially a hexadecimal hash string that is dependent on compilation inputs, particularly, the source files and compiler optimization settings. If these elements remain unchanged, the build ID will also remain consistent. This characteristic makes the build ID superior to a general-purpose file hash, which can vary due to non-essential metadata such as timestamps or debugging strings. The build ID can be directly extracted from a specific metadata section  \textit{.note.gnu.build-id} of a library file. It should be noted that build IDs alone are not adequate for establishing the relative order of the blobs (i.e., which blob has a more recent version).

% This method enables us to track changes in GPU drivers across historical firmware versions. 
\item Internal Blob Version:  
To overcome the Build ID's limitation, we introduce another component to our versioning scheme. We leverage the fact that certain GPU blobs generate debug logs during their initialization phase. These logs often include an internal version string with numerical components that can be compared. Specifically, Qualcomm GPU blobs use the format \verb|EV{major}.{minor}.{patch}|, while ARM GPU blobs employ a version string in the format of \verb|r{major}p{patch}|.
We can apply these regular expressions to retrieve the internal version numbers. However, it is worth noting that these debug strings may not always be present if the GPU blob developers enable aggressive optimizations during the build process.

\item LLVM Compiler Version: In the event of missing internal version numbers, we resort to another observation: both Qualcomm and ARM GPU blobs incorporate an LLVM compiler library.
The rationale behind GPU blobs embedding a compiler lies in their need to process shader programs. Specifically, shader programs are executed by the GPU for graphics rendering. They are typically written in a C-like high-level language such as GLSL or HLSL. As such, they must be compiled into corresponding GPU hardware instructions prior to execution. The LLVM library itself contains a version string following a format of \verb|{major}.{minor}.{patch}.{commithash}|. We employ a regular expression to search for this version string if the LLVM binary is not stripped. 
In case of a stripped library, we can still use fingerprinting strategies to estimate the LLVM library version.
One simple fingerprint is the textual strings found in the prevalent logging statements within the LLVM library. As LLVM evolves, for instance, with the introduction of new optimization passes, new strings are frequently added. This characteristic makes it a robust method for identifying the LLVM version. In addition, we can incorporate a more complicated fingerprint known as Binshape~\cite{shirani2017binshape}. This approach extracts a combination of features from the body of each function, such as the initial byte sequences (i.e., the prologue), call graphs, and statistics of machine instructions within a function. This approach allows us to identify not only the version but also the LLVM function names in a stripped binary.
\end{enumerate}

% It should be noted that the fallback technique may be insensitive to minor upgrades in LLVM compilers.

\subsection{GPU Blob Fuzzing}
The upcoming task it to identify potential bugs within the GPU vendor blobs. A commonly-used method is fuzzing, where we supply a system with randomly generated inputs and observe any resulting exceptions. For GPU, a particular fuzzing technique known as metamorphic testing is often favored~\cite{donaldson2019metamorphic}. This process commences with a reference shader program, typically procured from readily available mobile games. We then apply various transformations that preserve the semantics of the source code (e.g., converting a `for' loop to a `while' loop). This generates a collection of shader variants with heavily modified source code, yet maintaining the same output effect. If a variant shader produces an output significantly different from the reference shader, it indicates a potential bug in the GPU stack. This method is employed by 
the state-of-the-art GPU fuzzer \textit{GraphicsFuzz}~\cite{donaldson2020putting}. However, for our large-scale study, we find GraphicsFuzz unsuitable since it requires physical access to mobile devices. Furthermore, GraphicsFuzz provides a fuzzing speed of less than ten shaders per second. Considering the substantial volume of GPU blobs in our dataset, we require a fuzzer with a higher speed to complete the analysis within a practical timeframe.

\begin{figure}[t]
 % \vspace{-15pt}
  \begin{center}
  \includegraphics[width=1\linewidth]{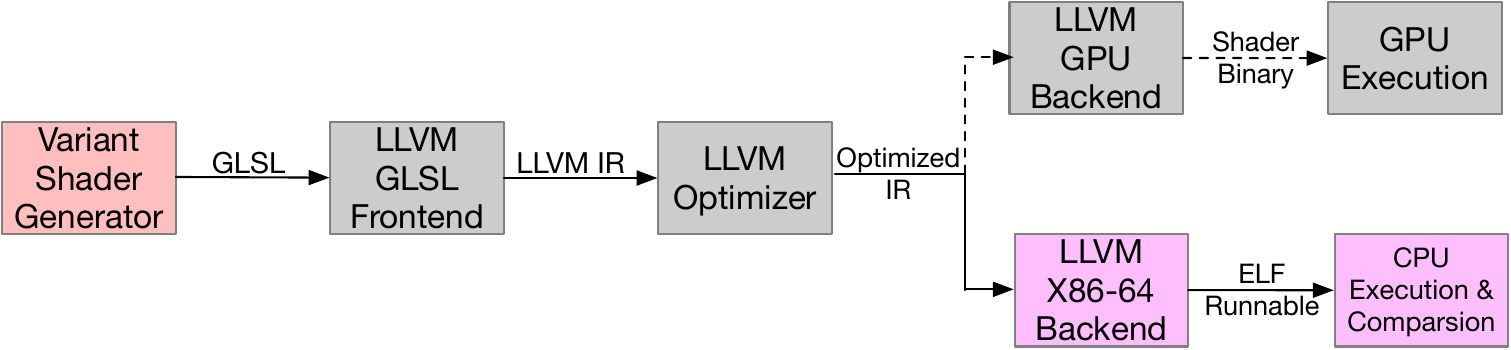}
 % \vspace{-10pt}
  \caption{LLVM Compiler Pipeline. }
  \label{fig:fuzzer}
  \end{center}

  \end{figure}

In this paper, we develop a performant offline fuzzer for the proprietary GPU blobs. We exploit the fact that both Qualcomm and ARM GPU blobs are constructed on the LLVM compiler infrastructure. LLVM encompasses a three-phase compilation process, as depicted in Figure~\ref{fig:fuzzer}. The frontend is responsible for converting the shader source code into an intermediate representation (IR). This IR is a platform-agnostic representation that preserves the semantic meaning of the source code. Subsequently, the optimizer ingests the frontend IR and applies a series of transformations to augment the code's efficiency. Exemplary transformations encompass constant folding, dead code elimination, and loop optimizations. Finally, the backend is responsible for translating the optimized LLVM IR into machine code for the target GPU architecture. 

Our fuzzer instruments the LLVM library to capture IR after the optimization passes. Instead of sending the IR to a GPU backend, we reroute it to an X86 backend to create an executable for native CPU execution. This approach lets us efficiently compare the output of the reference shader and variant shader without access to mobile phone GPUs. Below, we outline the implementation details of our fuzzer.

% Any discrepancy between the outputs suggests the presence of a bug, which can then be further verified on a physical device.
% If there is a discrepancy in the output, it suggests the presence of a bug in the first two phrases (i.e., the frontend pass or the optimization passes). 
% Since LLVM allows us inspect the IR output before and after each pass, we can precisely pinpoint the bug location. 
% In addition, we implement an interpreter to execute the resulting object files from the GPU backends. This allows us to identify potential bugs in the final stage in the compilation pipeline.
% In the following section, we will primarily focus on Qualcomm GPU drivers due to their substantial market share. Nevertheless, these techniques can be easily adapted to Mali GPU drivers with minimal adjustments.

\textbf{Reference Shader Database:}
Our fuzzer requires a reference shader database to generate shader variants. In this study, we curate a shader database from real-world games. We commence by crawling and analyzing the popular Android app stores such as \textit{Google Play}, \textit{APKPure}, and \textit{APKMirror}. As of Feb 2024, we compile a list of 145,732 gaming apps from these platforms. We also gather relevant metadata, such as game descriptions, categories, and download counts. Through our analysis on the metadata, we identified 10 popular game genres including 1) Action, 2) Adventure, 3) Board, 4) Causal, 5) Puzzle, 6) Racing, 7) Role-Playing, 8) Simulation, 9) Sports, and 10) Strategy. For each genre, we employ a weighted random sampling technique to select 20 games. The weight assigned to each candidate app is proportional to its download count. It ensures our selection leans towards popular games in the app stores. 
% We then proceed to extract shader programs from the selected games. 
% Notably, game binaries may compress, encrypt, and obfuscate shader programs into various data formats. This makes employing a static analysis approach to extract shader programs  challenging.
Afterward, we use Patrace\footnote{https://github.com/ARM-software/patrace} to extract the shader extraction from the selected games. Specifically, Patrace is designed to capture GLES calls at runtime from a gaming app and record them into a trace file for performance analysis. To extract the shader sources, we can filter out the \verb|glShaderSource| command from the trace files.
In our study, we assigned a researcher with extensive experience in mobile gaming to evaluate each selected game for a minimum of 10 minutes. 
% The evaluation was conducted on an ASUS ROG Phone 6, which is equipped with an 8-core CPU clocked at up to 3.2 GHz and 12 GB of RAM. 
During this process, we activate the Patrace's capturing functionality, which results in approximately 2000 minutes of trace recording. From these trace files, we recover 18,923 shader programs.
% One technical challenge we face is that Patrace requires setting the environment variable \verb|LD_LIBRARY_PATH| to Patrace's file path before launching a gaming app. This step instructs the Android dynamic linker to inject Patrace into the game's memory space. However, this method typically demands root access to the Android device, a privilege that is often restricted by most mobile vendors today.
% To circumvent this security measure, we employ the APK static patching technique to integrate Patrace directly into a gaming app. First, we use a tool called \textit{APKTool} to extract all files from an APK. Then, we inject a library dependency to Patrace into the game's native engine library using a tool called \textit{Patchelf}. Finally, all modified files are repackaged into a new APK and re-signed for installation.

\textbf{Variant Shader Generation:}
To generate the shader variants, we adapt a series of semantic-preserving transformations from Graphicfuzz on our shader database. These transformations can be categorized as follows: 
\begin{enumerate}
\item Statement Mutation: This transformation process takes a simple statement denoted as $origin$ and converts it into a more complex one using constructs from GLSL. One exemplary construct is the function mix($origin$, $unused$, $c$), which yields a linear combination of $origin * c + unused * (1 - c)$. Consequently, if we set $c = 1$, the output will remain as $origin$. 
% Another example involves the use of vector and index constructs, for instance, vec2($unused$, $origin$)[1] still yields $origin$.
\item Control Flow Mutation: In this transformation, we can interchange certain control flow constructs, for example, switching from `if' to `switch', or from `for' to `while'. We can also insert additional control flow elements, such as wrapping a code segment in a loop that executes only once. Alternatively, we can make a control flow construct more complex, for instance, unrolling a loop or splitting a loop. 
\item Code Donation: In this transform, we extract a group of statements from a randomly selected reference shader. These statements are then injected into the target shader. To avoid affecting the output of the target shader, we randomly rename each variable in the donated code and store the computational results to some unused built-in output variables.
\end{enumerate}
These transformations can be chained in a random order and in a recursive manner to generate complex variant shaders.

\textbf{LLVM IR Capturing:}
% We reroute the LLVM IR after the optimization stage to a x86 object generation backend. 
After generating shader variants, we feed them to the GPU blobs to perform shader compilation and capture the LLVM IR. 
One approach is to leverage the standard OpenGL graphics interfaces exposed by the GPU blobs. 
Specifically, we can first leverage the \verb|eglCreateContext| API to initialize a global EGL context, which serves as an interface between OpenGL APIs and underlying hardware. Afterward, 
a shader object can then be created using the \verb|glCreateShader| function. Subsequently, the shader source code is attached to the object using \verb|glShaderSource|. Finally, the shader can be compiled using \verb|glCompileShader|. To capture the LLVM IR during the compilation, we can leverage a LLVM command line option, \verb|-print-after-all|. It instructs the compiler to print out the IR after each optimization pass for debugging purposes. We can inject this option by invoking an LLVM function named \verb|cl::ParseCommandLineOptions|. The memory address of this function can be identified using the string fingerprint method, as discussed in Section~\ref{versionid}.

% Upon successful compilation, multiple shader objects can be further attached to a program object using the \verb|glAttachShader| function, followed by their linkage using \verb|glLinkProgram|. A program amalgamates multiple shader objects to define a full rendering pipeline. 
To execute the aforementioned APIs without access to physical devices, 
we can set up a QEMU full-system emulator to run a vanilla Android system (i.e., AOSP). By default, the emulated Android system employs a software-based OpenGL implementation. We can upload the proprietary GPU blobs to the emulator and replace the default implementation by adjusting an environment variable named \verb|debug.gles.layers|\footnote{https://developer.android.com/ndk/guides/rootless-debug-gles}. 
This action alone is not sufficient. When creating an EGL context, the proprietary binary needs to issue IOCTL system calls to the GPU kernel module for querying hardware information such as GPU model and VRAM size. Since this kernel module is absent in the emulator, an EGL exception will be raised.
We circumvent this issue by implementing a shim Linux kernel module for the emulator. We stub out most IOCTL calls in the original GPU kernel drivers and retain only those necessary for EGL context creation.

% The driver is adapted from the original Qualcomm kernel mode driver called MSM. We stub out most IOCTL calls, which manipulate underlying hardware registers, to execute nop operations. We only retain a minimum number of IOCTL calls to allow the proprietary GPU driver to create an EGL context and conduct shader compilation, such as hardware model query and memory allocation. 

For Qualcomm GPU blobs, we can adopt a more efficient approach by exploiting the fact that Qualcomm GPU drivers partition the compiler component into a separate library named \textit{libllvm-glnext.so}. This library relies only on C runtime libraries (bionic C) and exposes accessible compilation interfaces. This allows us to directly invoke the compiler component via a user-space ARM CPU emulator (i.e., \verb|aarch-qemu|), which is more performant than full system emulation.
% This is more performant than emulating the full Android system. 
Specifically, we first utilize the dynamic linking API \verb|dlopen()| to load the target library into the memory. Afterward, we retrieve the function pointers of compiler interfaces via the API \verb|dlsym()|.  A primary function is \verb|QGLCCompileToIRShader|, which ingests a shader source code string buffer and returns the optimized LLVM IR.  An illustrative snippet of LLVM IR is presented in Figure~\ref{llvmsnip}.
% The primary function is \verb|QGLCCompileToIRShader| and \verb|QGLCLinkProgram|.
% The former function takes a shader source code string buffer and compile them into optimized LLVM IR, while the latter function further takes the optimized LLVM IR from different shaders and link them into a single GPU program object. 
% It should be noted that QEMU userspace emulation requires an ELF interpreter/linker and libc runtime libraries. They can be obtained from the Android system partition.

\begin{figure}
\begin{minipage}{0.5\textwidth}
\begin{lstlisting}[language=LLVM, frame=single]
; Input/Output Variables 
@a_color = external global <4 x float>, align 16
@v_fragmentColor = external global <4 x float>, align 16
define void @llvm_main() {
  ; Shader Input
  %reg_a_color_0 = call float @llvm.qgpu.fget.reg.f32.p0v4f32
  (<4 x float>* @a_color, i32 0, i32 1)
  ; Shader Math Operations
  %color_rsq = call float @llvm.qgpu.rsq(float %reg_a_color_0)
  ; Shader Sample Functions
  %texture_val = call <4 x float> @llvm.qgpu.fsampler.v4f32.
  v2i16.v2f32.i32 (i32 0, <2 x i16> zeroinitializer, 
  <2 x float> %sample_coords, i32 undef, 
  <4 x i32> <i32 128, i32 0, i32 0, i32 0>, 
  i16 0, i16 0)
  ; Shader Output
  call void @llvm.qgpu.global.fset.reg.p0f32.v4f32
  (float* getelementptr inbounds (<4 x float>* 
  @v_fragmentColor, i32 0, i32 0), <4 x float> %4,i32 0,i32 4)
}
\end{lstlisting}
\end{minipage}
\caption{Example LLVM IR Code for an OpenGL Shader}\label{llvmsnip}
 \vspace{-10pt}
\end{figure}

\textbf{CPU Binary Generation and Execution}
The next step is to generate a CPU executable from the captured IR. Specifically, we first use the LLVM static compiler \textit{llc} to translate the IR into x86 assembly language. The resulting assembly output is then passed through the x86 assembler \textit{llvm-mc} and the x86 linker \textit{lld} to produce a native executable. 
Special attention is required during the linkage phase, as the captured IR often includes invocations to various intrinsic functions. These functions represent computational operations that can be translated into efficient GPU machine instructions. They are identifiable by their unique naming convention, which begins with \textit{llvm}. Since these intrinsic functions are not available on the X86 platform, we provide our software implementation to prevent linkage errors as follows.

\textit{Shader Input/Output}: 
    % One category of intrinsic functions manages access to input and output variables in shader programs.
    In OpenGL, an input variable and output variable can be defined using the keywords \verb|in| and \verb|out|, respectively.  
Access to these variables is translated to intrinsic function invocations, \verb|llvm.qgpu.fget| and \verb|llvm.qgpu.fset| respectively. These functions accept a GPU memory address as input. In our backend, we remap the GPU addresses into heap memory regions, which are pre-allocated using the system call \verb|mmap|. At the entry point of each shader program, we populate the input memory regions with randomly generated values using predetermined seeds. Upon the shader program's exit, we generate hashes for the contents of the output memory regions. This allows us to compare the output of a variant shader with that of a reference shader.

% OpenGL also supports uniform variables, which provides lower-latency access and can be declared with the keyword uniform.

\textit{Math Operations}: Another common type of intrinsics encompasses mathematical computations. For example, the intrinsic \verb|llvm.qgpu.rsqf| is employed to compute the reciprocal square root of a floating-point number.
% a calculation heavily used in graphic rendering. 
In a Qualcomm GPU, this intrinsic can be directly mapped into the machine instruction \verb|rsq|, which consumes only three GPU cycles. In our X86 backend, we need to remap it to corresponding math functions within the \verb|libm.so| library.
A technical challenge is that OpenGL allows developers to use half-precision floating point numbers in a shader.
However, on the X86 platform, the libm library does not support half floating-point computations. One naive workaround is to emulate half-precision operations using software (e.g., Berkeley Softfloat). This inevitably incurs significant performance penalty. Instead, we implement an LLVM transformation pass that iterates over all IR instructions and examines each float operand. If an operand is of half-precision, we then promote it to single precision. Special attention is given to the \verb|fpext| instruction, which extends a value from a smaller to a larger floating-point type. Since all floating-point operands have already been promoted to the highest precision, we convert \verb|fpext| to \verb|bitcast|  to make it a no-op operation. 
This strategy is implemented at compilation time and does not incur any runtime performance overhead.

\textit{Shader Sampler}: Another common type of intrinsic functions is sampler function, for instance, \verb|llvm.qgpu.fsampler|. A sampler is used to fetch and process texels from texture resources. It controls various aspects of texture representation, such as filtering, wrapping, and mipmapping. Executing these sampler functions on x86 CPUs can be computationally intensive and inefficient. Instead, we simplify the sampler function as a hash function. Given the same sampling parameters and global random seed, it consistently generates the same random texture output. It behaves as if we are providing procedurally generated textures to the shaders.

We implement these intrinsic functions in C and compile them into object files using Clang. These object files are subsequently linked with the LLVM IR to produce an ELF executable. In this process, we need to address differences in the Application Binary Interface (ABI), which governs the data exchange protocol between object files originating from different source languages.
Specifically, when LLVM attempts to pass a small float vector (e.g., \verb|4 * float|) to C, the data is, by default, passed through an xmm register. However, a C function expects the float array to be passed on the stack. To resolve this issue, we explicitly define the incoming parameter in the C function using Clang's \verb|ext_vector_type| attribute. This enforces Clang to retrieve incoming float arrays from the xmm register. 
% Additionally, when LLVM passes two-element 32-bit integer vectors (e.g., \verb|<2 * i32>|) to C, the two integers are placed on the 64-bit boundary inside an xmm register, instead of 32. This means that on the C side, to fetch the second element, we need to use index 2 instead of 1 in the array.

\section{Evaluation}
In this section, we present our research findings for our two primary research questions: 1) What is the update pattern of GPU vendor blobs? and 2) How vulnerable are these GPU vendor blobs? Additionally, we conduct a case study to exploit these vulnerabilities to launch a denial-of-service or even arbitrary code execution attack on the target device.

% questions:
% \begin{itemize}
%  \item \textbf{RQ1 (Driver Version)}: What versions of GPU drivers engines are being utilized? 
%  \item \textbf{RQ2 (Update Frequency)}: Do the vendor GPU drivers receive security updates? If so, what is the update frequency?
%  \item \textbf{RQ3 (Update Differences)}: Do update frequencies vary among vendors or across different devices from the same vendor?
%  \item \textbf{RQ4 (Vulnerability Impacts)}: To what extent are the vendor GPU engines affected by behavior bugs and security vulnerabilities? 
% \end{itemize}
\subsection{Experiment Setup}
We deploy our \textit{GPUBlob-Inspector} pipeline on an Ubuntu 24.04 machine. The machine is equipped with two AMD EPYC 7742 64-Core Processor CPUs, 1024 GB of RAM, 128 TB SSD hard disks, and a 10 Gbps network connection. We provide a detailed breakdown of the time taken at each stage of the pipeline.
The initial phase of firmware crawling (approximately 13,000 images) is completed within 27 hours. The subsequent task of firmware extraction is accomplished in 147 minutes. The average processing time for each firmware is approximately 38 seconds, with the maximum processing time recorded at 278 seconds.
The version identification stage is completed within 22 minutes, with an average processing time of 12 seconds and a maximum of 19 seconds per GPU blob. 
In the fuzzing stage, we generate 200 variants for each reference shader. The generation process takes approximately 491 minutes.
We then input the variant shaders into a GPU blob to generate LLVM IR, which consumes 189 minutes. The  compilation of these IRs into x86-64 executable ELF files requires 127 minutes. Executing these files and comparing their outputs takes about 7 minutes.
It is important to note that the duration of the fuzzing stage is contingent on the number of shader variants generated. Increasing the number of shader variants enhances the probability of bug discovery but also prolongs the process.

\subsection{RQ1: Driver Update Frequency}
Our pipeline successfully analyzed 13,901 firmware images. Among these, a significant proportion, 78\%, contain Qualcomm GPU blobs. Only 21\% of the images contain ARM GPU blobs. The remaining fraction is made up of Nvidia Tegra and MediaTek Dimensity. Due to their minimal market shares, we exclude these two from our results.
We further de-duplicate all the GPU blobs in our dataset using their Build IDs.
We identify 792 distinct versions. Specifically, 65\% are provided by Qualcomm, while 35\% are sourced from ARM. 
We also provide the GPU composition ratio across several top-tier mobile device manufacturers in Table~\ref{tab:gpucomp}.
The device count also indicates a significantly higher popularity of Qualcomm GPUs compared to ARM GPUs. We  observe that Qualcomm GPUs are predominantly found in mid-range to high-end devices potentially due to their higher graphics rendering capabilities. In contrast, ARM GPUs are typically utilized in entry-level models. These findings are consistent with a previous report\footnote{https://www.techcenturion.com/mobile-gpu-rankings}.

% We further de-duplicate all the GPU blobs in our dataset using their Build ID.
% We identify 792 distinct versions; 65\% are provided by Qualcomm, while 35\% are sourced from ARM. 
% which indicate that Qualcomm, due to its prevalence in mid-range and high-end devices, holds a larger market share, while ARM is more commonly found in entry-level devices. This observation aligns with prior market reports,

% \begin{figure}[t]
%  % \vspace{-15pt} 
% \begin{center}
% \includegraphics[width=1\linewidth]{figures/driver_count.pdf}

%   \caption{GPU Composition Across Major Manufacturers}
%   \label{fig:devicedistribution}
%   \end{center}

% \end{figure}

\begin{table}[]
\caption{GPU Composition Across Major Manufacturers}
\label{tab:gpucomp}
\centering
\begin{tabular}{|l|l|l|l|}

\hline
Qualcomm / ARM & Firmware & \begin{tabular}[c]{@{}l@{}}Distinct\\ Driver\end{tabular} & Device \\ \hline
Samsung & 73\% / 27\% & 60\% / 40\% & 61\% / 39\% \\ \hline
Xiaomi  & 70\% / 30\% & 67\% / 33\% & 73\% / 27\% \\ \hline
Oppo    & 72\% / 28\% & 64\% / 36\% & 62\% / 38\% \\ \hline
Google  & 75\% / 25\% & 56\% / 44\% & 58\% / 42\% \\ \hline
OnePlus & 82\% / 18\% & 61\% / 39\% & 78\% / 22\% \\ \hline
Total   & 78\% / 22\% & 60\% / 40\% & 69\% / 31\% \\ \hline
\end{tabular}
\end{table}

\begin{figure}[t]
 % \vspace{-15pt} 
\begin{center}
\includegraphics[width=0.95\linewidth]{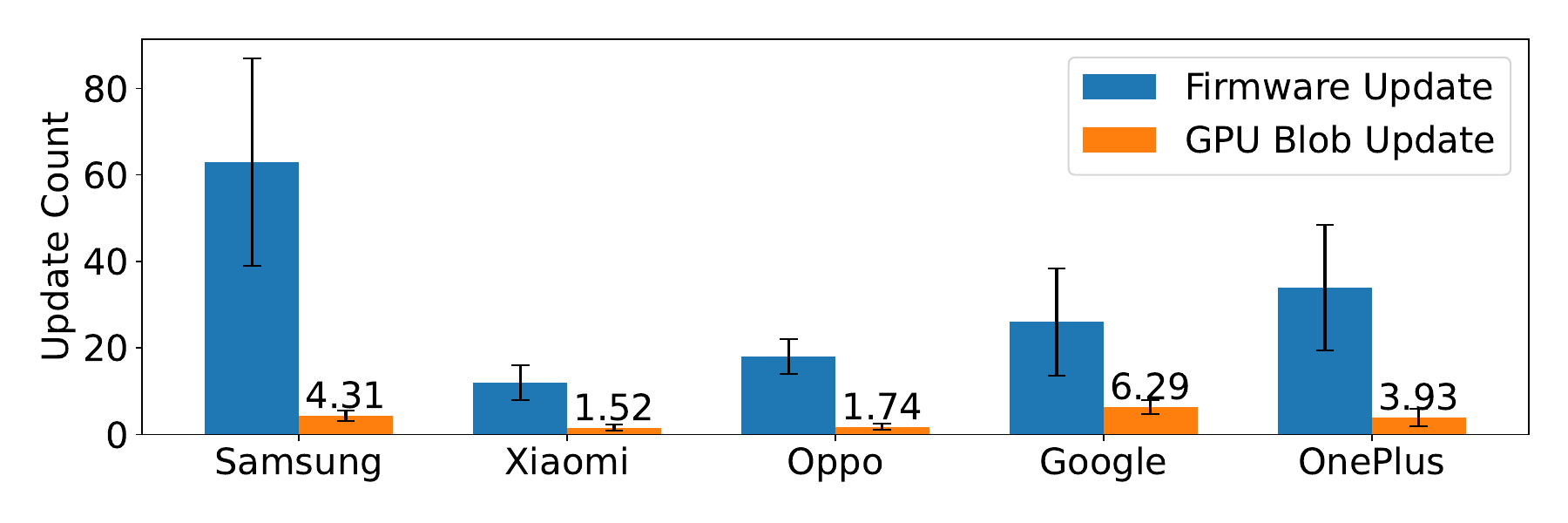}

  \caption{Update Count for Firmware and GPU Blobs per Device}
  \label{fig:deviceupdatecount}
  \end{center}
\end{figure}

\textbf{Vendor Blob Update Count per Device}: We quantify the number of firmware and GPU driver updates across different manufacturers. The findings are presented in Figure~\ref{fig:deviceupdatecount}. For comparative purposes, the average update number per device is reported. A notable observation is that the number of firmware updates substantially surpasses that of GPU vendor blobs. It suggests that device manufacturers tend to prioritize firmware updates in a device's life cycle. 
Another interesting observation is that
certain manufacturers, such as Xiaomi and Oppo, exhibit an average GPU driver update count close to one. This indicates that the majority of their devices rarely receive GPU vendor blob updates throughout their lifecycle. In stark contrast, Google devices receive, on average, six GPU vendor blob updates. This discrepancy could be attributed to Google's role as the first-party developer of the Android operating system. As such, Google likely has more resources for device maintenance and possesses greater influence over its software and hardware ecosystem.

% For those devices that do receive GPU blob updates, we further illustrate the intervals between each updates across mobile phone manufacture in Figure~\ref{fig:deviceinterval}. We can also observe notable differences between vendors; Xiaomi and Oppo receive updates with a frequency of less than once per year, on average 371 days, while Google typically receives a GPU driver update approximately every 67 days. 

% \begin{figure}[t]
%  % \vspace{-15pt} 
% \begin{center}
% \includegraphics[width=1\linewidth]{figures/update_interval.pdf}

%   \caption{Update Interval for System and GPU Drivers}
%   \label{fig:deviceinterval}
%   \end{center}

% \end{figure}

\textbf{Vendor Blob Update Delay: } 
A blob update delay is defined as a situation where a firmware release contains a GPU blob version that is not as up-to-date as the version already in use by another device with the same GPU hardware. This means the firmware release could have potentially utilized the newer blob for enhanced performance or security.
The degree of the GPU blob update delay can be estimated as follows:
\begin{enumerate}
\item Given a target firmware image with a release date $r_f$ and its associated GPU blob version $v_o$, we search our firmware database for the oldest firmware image containing the same blob and denote its release date as $r_o$. This process estimates the date when this version of the GPU blob was first introduced.
\item Following this, we filter our database for firmware that is equipped with the same GPU model and is released prior to the date $r_f$. Among these filtered results, we identify the most recent GPU driver version, $v_l$. In other words, the target firmware could have updated the blob version from $v_o$ to $v_l$.
\item In the final step, we search our firmware database for the earliest firmware that incorporates the driver version $v_l$ and denote its release date as $r_l$. We then can compute   the delay as $D=r_l-r_o$, which estimates the degree of blob outdatedness. 
\end{enumerate}

\begin{figure}[t]
 % \vspace{-15pt} 
\begin{center}
\includegraphics[width=0.90\linewidth]{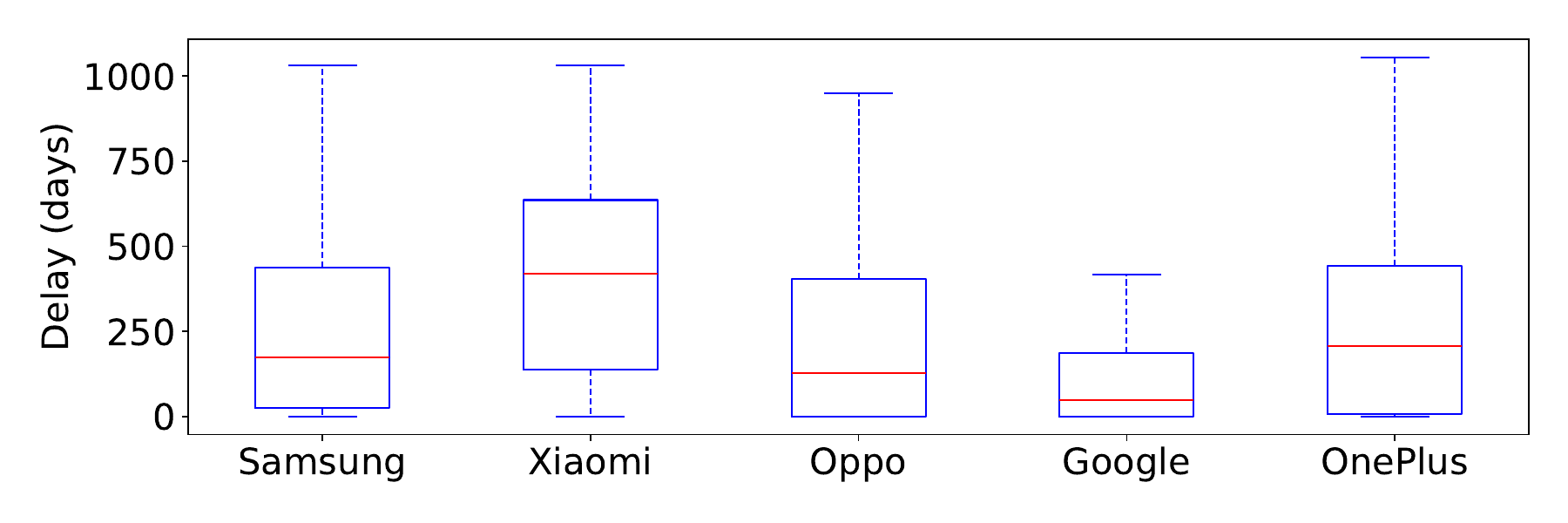}

  \caption{Blob Update Delay Estimation}
  \label{fig:blob_delay}
  \end{center}

\end{figure}

From our dataset, we observed that 82\% of firmware images contain outdated GPU blobs. The median value of the delay $D$  is $273$ days, with the highest value reaching $1,128$ days. This observation suggests that most device manufacturers neglect timely updates on  vendor blobs, potentially leaving devices in a more vulnerable state.
In Figure~\ref{fig:blob_delay}, we provide a detailed breakdown of the results based on device manufacturers. 
We can infer that certain manufacturers, such as Google, update their proprietary GPU drivers significantly more promptly than others.

We further investigated whether the update delay is influenced by the GPU vendors. Our research indicates that 85\% of firmware images utilizing Qualcomm GPUs contain outdated GPU blobs. This is in contrast to the 64\% of firmware images with ARM GPUs that are outdated. When considering the median delay for firmware updates, Qualcomm devices lag behind at 281 days, while ARM devices show a slightly better result at 231 days. It is concerning to note that Qualcomm devices, despite their significantly larger adoption rate, tend to receive blob updates less promptly.

\begin{figure}[t]
 % \vspace{-15pt} 
\begin{center}
\includegraphics[width=0.90\linewidth]{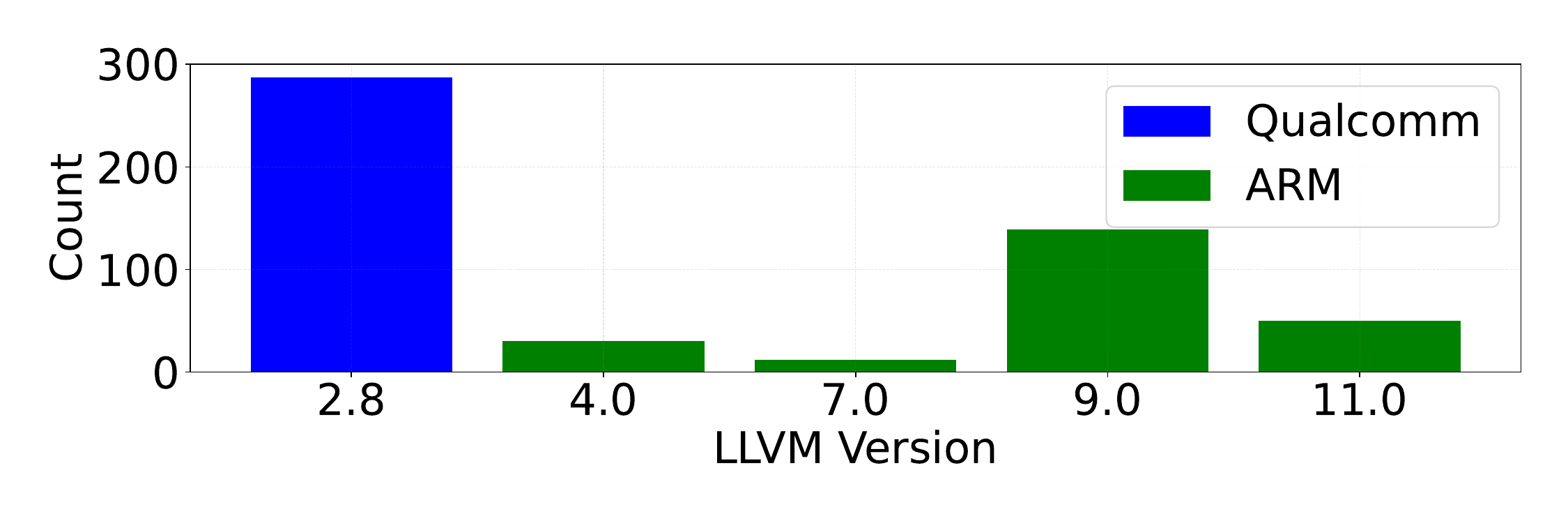}

  \caption{LLVM Version Distribution}
  \label{fig:llvm_ver}
  \end{center}
\end{figure}

\textbf{LLVM Compiler Outdatedness:} A key element of the GPU blobs is the LLVM library. We investigate their LLVM versions and present their distribution in Figure~\ref{fig:llvm_ver}. An important observation is that all Qualcomm GPU blobs rely on the considerably outdated LLVM version 2.8, which is released nearly 15 years ago. In contrast, ARM appears to adopt more recent LLVM versions. Specifically, 
the majority of ARM blobs are based on LLVM 9.0, followed by LLVM 10.0, and LLVM 11.0. The remaining drivers employ LLVM 5.0 versions. However, these versions are also relatively outdated, with LLVM 11.0 having been released in 2020.
This implies that all the GPU blobs could potentially be susceptible to a multitude of LLVM bugs that have been discovered since their respective release dates.
One potential reason for Qualcomm and ARM's reluctance to update their LLVM library could stem from the LLVM project's emphasis on innovation and performance enhancements. This focus often results in significant API changes across different LLVM versions and managing these breaking changes is known to be a challenging and labor-intensive task.
  
% \begin{center} 
% \fbox{

%   \parbox{0.45\textwidth}{ % Replace \textwidth with your desired width
%     Answer to RQ1: Mobile vendors infrequently update their proprietary GPU driver binaries after the initial release of their devices. The driver update frequency is notably less than that of Android operating system updates.
%   }
% }
% \end{center}

\subsection{Vulnerability Impacts}
In this section, we present the fuzzing outcome for the most recent Qualcomm GPU blob to highlight  potential security vulnerabilities. This particular blob is sourced from a Xiaomi \textit{Note 13 Pro 5G} device firmware, released on February 23, 2024, with the internal blob version number \verb|031.42.23.11|.

\textbf{Vulnerabilities Finding:} Over the course of approximately 14 hours of fuzzing, our fuzzer identifies 289 instances of irregular program behaviors or incorrect program outputs.
To ensure that the anomalies identified are not a result of a bug within our fuzzer,
we validate these instances on physical devices. In brief, we substitute the reference shader in the retrace file with a variant shader. Following this, we employ Patrace to replay the updated retrace file. 
We utilize an offscreen rendering mode, which facilitates us to transfer the rendered frames into main memory and store them onto a hard disk file.
Subsequently, we compare the frames with the reference frames using the chi-square test on their respective histograms. If the test indicates a significant difference, an offending instance is validated. 
We perform a thorough analysis on the offending variants shaders to pinpoint the root causes of anomalous behaviors.
We categorize the causes into several distinct groups as follows.
% A KL distance of zero implies no differences  between the two frames. Conversely, if the frames show significant differences, the KL value can be quite large (e.g., greater than 0.5). 

% To accelerate this process, we employ  two analysis techniques. The first useful approach is to  utilize the LLVM debugging command line option \verb|-print-after-all| to output the IR of each optimization pass. By compiling various IR versions and monitoring which ones display incorrect behaviors,  we can pinpoint the optimization passes that result in errors.
% The second one is shader reduction. When a bug is revealed by a variant shader, the shader code is often lengthy due to numerous steps of semantics-preserving transformations. However, only a small portion of this code might be necessary to activate the bug. The reduction process reverses these transformation steps to pinpoint the specific change that triggers the bug. 

% One useful approach is to perform reducing operations on a variant shader. When a variant shader exposes a bug, it is typically very large due to the aggressive semantics-preserving transformations. However, only a fraction of this code may be needed to trigger the bug. The reducing operation reverses those transformations step by step, to pinpoint the transformation that triggers the bug originates.

\textit{1. Memory Access Violation: }
The first type of offending variants triggers memory access violation during the shader compilation stage (i.e., during the invocation of \verb|glCompileShader|). This results in an immediate crash of the application process. 
We conduct an analysis of the 29 problematic cases using their stack traces. Our observations reveal that 5 of these cases are triggered by null pointer references, while the rest result from invalid memory access violation. Upon further investigation of these invalid memory addresses, we find that most of them are low memory addresses, such as 0x7 or 0x1b. These could potentially be produced by an offset operation with a zero base address, which once again ties back to a null pointer error. However, we identify four addresses as heap memory. By instrumenting the application process with the memory checker Valgrind, we infer that these are likely triggered by a use-after-free heap error. This is alarming, as it potentially allows an attacker to construct arbitrary code execution exploits. 
Additionally, we aim to identify the positions of the bugs. One useful technique is to utilize the LLVM debugging command line option \verb|-print-after-all|, which displays the  output IR of each LLVM component. If a component fails to generate output IR, we can identify it as the faulty one.
Our findings reveal that 7 of them are found in LLVM analysis passes, including type-based alias analysis and dead code analysis. Another 17 originate from LLVM function optimization passes, such as peephole optimization, dead code elimination, and constant expression folding. The remaining 5 are located in backend optimization passes, specifically machine code constant folding and peephole optimization.

\textit{2. Compilation Stalls: } The second type of problematic shader variants results in the application process stalling during the  shader compilation. Our fuzzer identifies three such instances. By attaching a GNU GDB debugger to the compiler thread and performing single-step execution, we observe that they are induced by infinite iteration of  optimization passes, such as loop unrolling, control flow graph simplification, and instruction combination. In LLVM, the execution of an optimization pass generates a new set of LLVM IR, which may open up additional opportunities for optimization. To harness them, certain optimization passes are designed to operate iteratively. However, it’s not unusual for logic flaws to be present within these passes and exemplary bug reports can be found in \cite{bug1,bug3,bug2}. They prevent the IR from achieving a stable state, resulting in the optimization pass being executed repeatedly.
As such, a malicious actor could potentially craft a denial-of-service shader payload, which can cause the targeted application to become unresponsive. 

% \textit{3. Assertion Failures:}
% Assertions are written by developers in the source code to check for preconditions expected by the code following it.

\textit{3. Incorrect Program Semantics:}
The most common type of offending shader variants causes the compiler to produce incorrect program semantics. This kind of errors may cause artifacts in the graphics output and does not necessarily lead to security vulnerability. Nevertheless, as suggested by prior research~\cite{alves2016software}, we anticipate a positive correlation between the number of these behavior bugs and the count of security vulnerabilities.
To identify the incorrect section of program semantics, we can generate and compare Data Dependency Graphs (DDGs) of the variant IR and the reference IR.
A DDG is a directed graph where nodes represent instructions, and edges denote data dependencies between these instructions. For example, if a variable is defined in instruction A and used in instruction B, there will be an edge flowing from A to B. Given a DDG, we start from instructions that produce shader output variables and backtrack the graph until we reach all instructions that consume the shader input variables. By doing so, we can identify the chain of instructions used to transform those input variables into an output variable. For a correct variant program, its instruction chain should be semantically equivalent to that of the reference program.
Using DDGs, we pinpoint several flawed optimization passes, including instruction combination, dead code elimination, and loop simplification. For example, when dealing with a complicated control flow, these passes may incorrectly deduce that certain instructions do not affect the program's observable behavior.  If such an instruction is removed, and the values it defined are still used elsewhere in the program, those values are replaced with a keyword \textit{undef} (e.g., \verb|%2 = fmul float %1, undef|). According to the LLVM IR specification, the keyword undef has special semantics; it serves as a placeholder for a constant value that can be any arbitrary value. The compiler can replace undef with any value (e.g., zero or one) in a completely nondeterministic manner, leading to unpredictable shader output.
% A common discrepancy lies in the floating-point operation instructions. 
% Specifically, we observed that 17 offending shaders contain floating-point instructions with undefined values as parameters (e.g., \verb|%2 = fmul float %1, undef|). 

% We analyaze the passes that generate the undef values, they are Dead Code Elimination removes instructions that do not affect the program's observable behavior. When an instruction is removed, the values it defined can be replaced with undef if those values are used elsewhere in the program but are no longer relevant. and InstCombine, If an instruction's result is not used in a meaningful way, it can be replaced with undef
\textbf{Fuzzing Speed:} We conduct a performance comparison of our fuzzer against the state-of-the-art GraphicsFuzz. GraphicsFuzz features a client-server architecture. We run its server program \textit{glsl-server} on a desktop PC equipped with an AMD 5990x CPU and 128 GB of RAM for variant shader generation. Its client-side program \textit{gles-worker} is run on a high-end ASUS ROG Phone 8 to fetch, compile, and execute these shaders. GraphicsFuzz achieves a fuzzing speed of approximately 7 shaders per second. In contrast, our fuzzer operates at a significantly faster speed, processing about 75 shaders per second. This enhanced speed enables us to uncover more vulnerabilities within the same timeframe.

\textbf{Different Blob Versions:} We also validate the problematic shaders across 10 versions of GPU blobs, which are evenly sampled from the dataset based on time intervals. 
Our findings indicate that these problematic shaders consistently trigger the same anomalous behaviors in all selected GPU blobs, suggesting that most vulnerabilities have existed for a considerable period.
This aligns with the observation that Qualcomm GPU blobs have been using a very deprecated LLVM compiler for an extended period. 
% A shader program typically takes a set of global variables as input and writes the output to another set of global variables. 

% This appears to be the case in the Qualcomm GPU backend, where all undefined values are explicitly converted to zero before being translated into native GPU instructions. However, this is not the case in the X86-64 backend, where undefined values can sometimes be converted into other constant values to achieve more aggressive optimization passes, such as dead code elimination. To suppress this issue, We implement an LLVM function pass that iterates over all instructions in the function and checks each constant operand. If an operand is an undef value, it replaces it with a zero constant of the same type. 

\subsection{Case Study: WebGL}
In the previous section, we identified shaders that can induce insecure behaviors during compilation. This section presents a case study to exploit them, which may result in denial of service (via compiler stalling) or arbitrary code execution (via use-after-free memory violations) on a target device.

The first step is to identify a widely accessible interface that can accept arbitrary shader input for compilation. One such interface is WebGL, a JavaScript API for rendering 3D graphics in a web environment. WebGL is derived from the OpenGL ES specification and can be considered a subset of OpenGL ES. In WebGL, shaders can be compiled using the \verb|gl.compileShader| JavaScript interface. Internally, the shader programs are still processed by the vendor GPU blob, thus triggering its vulnerabilities. WebGL are not only available in system browsers but also in apps that embed browser engines (e.g., WebKit or V8) to support in-app browsing. These apps are not uncommon and they are generally referred to hybrid apps. According to a recent empirical study~\cite{wen2024keep}, 149 apps among the top 500 most downloaded apps are hybrid. 

To launch an attack via WebGL, an adversary can host a web page with malicious shader code on the internet. Users are then deceived into visiting this page in a system browser or a hybrid app through phishing emails, instant messages, or rogue wireless access point attacks.
The exploit shader is then downloaded and executed automatically on the victim's machine.
It should be noted that Chrome introduces an intermediary layer called \textit{Almost Native Graphics Layer Engine} (ANGLE)\footnote{https://github.com/google/angle} to optimize and sanitize input shaders before feeding them to vendor GPU blob. As a result, 
it necessitates the modification of our fuzzing pipeline to identify new offending shaders. Specifically, we incorporate ANGLE to sanitize generated variant shaders before feeding them into the LLVM compiler. In the course of a 4-hour fuzzing session, we identify 9 sanitized shaders that crash the Chrome browser.

\section{Discussion}
\textbf{GPU backend bug detection:} Our fuzzer redirects a shader program’s LLVM IRs to an x86 backend to generate a CPU executable. It allows us to compare shader outputs without access to mobile GPUs. However, this method bypasses certain GPU backend processes (e.g., instruction scheduling or register allocation). Therefore, our fuzzer is not capable to detect vulnerabilities in these stages. This implies that the data reported in our evaluation section represents a conservative estimate of the severity.
% If we were to consider the backend passes, the severity would be even more significant.
We are currently implementing an interpreter to  directly execute GPU instruction output from the GPU blobs using CPUs. This allows us to uncover more bugs in the GPU backend stages. 
Our interpreter adopts a traditional decode-execute loop design; upon decoding a GPU instruction, the interpreter dispatches it to a corresponding function for execution.  To facilitate the instruction decoding, we employ GPU disassemblers from the Mesa 3D Graphics Library\footnote{https://www.mesa3d.org/}. Before executing the program, 
we randomize all register values before a GPU program begins. Upon the program's completion, we compare the values in the register sets.  Any discrepancy could indicate a potential bug.

\textbf{Fuzzer Support for Vulkan and ARM:}
In this study, our primary focus is on the OpenGL ES graphics APIs. This is due to the fact that Android mandates each device manufacturer to provide an implementation of OpenGL ES. Recently, a number of high-end mobile devices have begun supporting Vulkan Graphics APIs. Compared to OpenGL ES, Vulkan offers reduced CPU overhead and more detailed control over GPU resources. To support Vulkan, the GPU vendors  integrate an extra LLVM frontend to transform Vulkan code (i.e.,  SPIR-V) into LLVM IR. 
As such, our fuzzer can also be adapted to support Vulkan by providing an Vulkan variant shader generator. 
In addition, this study places emphasis on Qualcomm GPUs, primarily due to its dominant market share. Nevertheless, our methodology can be easily replicated to ARM GPUs with little modification. Our preliminary experiments identifies 19 vulnerabilities in Mali GPU blobs. We speculate that ARM GPUs, benefiting from a more recent LLVM compiler version, may have fewer vulnerabilities compared to Qualcomm. Further investigation is warranted.

\textbf{Vulnerabilities in Vendor Blobs Beyond GPUs:}
We carry out preliminary measurements on other types of vendor blobs in Android firmware, such as accelerometers and fingerprint readers. Unlike GPUs, these blobs are only loaded into a single process, making them likely less critical to system security due to fewer exploitation code paths. They are also sourced from a wide range of hardware manufacturers, which complicates the comparison of update patterns and vulnerability discovery. Nevertheless, further research in this area is warranted.

\section{Related Work}
The Android firmware encompasses a variety of software components.
Extensive research efforts have been made to scrutinize their security implications.
% typically including the Linux Kernel, AOSP system framework, vendor drivers, and pre-installed apps.
% Extensive research efforts have been made to investigate the security implication of various Android components.

\textbf{Linux Kernel:} Linux Kernel is often identified as the primary source of vulnerabilities that can compromise Android system security. Consequently, it is subject to extensive scrutiny. For instance, Zhou et al.~\cite{zhou2014peril} conducted research investigating the security risks arising from hardware vendors' unsafe customizations in Linux kernel drivers. Several surveys~\cite{lu2019kernel,mazuera2019android} also indicate that a significant portion of Android vulnerabilities are located in kernel-mode drivers.
However, given that the kernel component is open source, it facilitates developers in conducting regular code reviews and swift integration of patches to mitigate vulnerabilities.
Zhang et al.\cite{zhang2021investigation} examined the Android kernel patch process, uncovering bottlenecks in patch propagation. Similarly, Farhang et al.\cite{farhang2019hey} investigated the latency between a hardware vendor releasing a patch and its integration into the Android kernel repositories. Recently, several studies~\cite{yu2021sepal,hernandez2020bigmac} also explore the kernel-level security module \textit{SEAndroid} to confine system services and reduce the kernel attack surface.

\textbf{AOSP System Framework:} 
% The framework component provides apps with access to system services (e.g., camera and location), the Java language runtime (i.e., ART), and low-level native libraries (e.g., SSL or OpenSSL). 
A considerable number of Android vulnerabilities also stem from the system framework~\cite{mazuera2019android}. Although Google frequently releases framework security patches to address these vulnerabilities, it is the responsibility of device manufacturers (OEMs) to distribute these patches to their users. Extensive research studies have examined this process. For instance, Zheng et al.~\cite{zheng2014droidray} developed a tool called \textit{DroidRay} to assess the security patch level of the Android system framework from various firmware images. Similarly, Hou et al.~\cite{hou2023can} evaluated a large-scale firmware dataset and found that even when a device claims to be updated to the latest system framework, there is no guarantee that all corresponding patches have been integrated by the manufacturers. Additionally, Jones et al.~\cite{jones2020deploying} conducted an extensive study to reveal that the Android framework security updates rollout process is effectively affected by the carrier-manufacturer relationship.
A recent study~\cite{brant2022study} also observes that the framework patches may not undergo thorough testing before being rolled out by manufacturers, thus resulting in low code coverage.

\textbf{Pre-installed Android Apps:}
Pre-installed apps often come with pre-approved, highly sensitive permissions and capabilities. If these apps contain vulnerabilities, they can significantly impact system security. Consequently, extensive research has been conducted to analyze vulnerabilities in these apps.
Gamba et al.~\cite{gamba2020analysis} analyzed pre-installed Android apps to understand their potential impact on device users such as personally identifiable information leakage and pervasive user behavior tracking. Elsabagh et al.~\cite{elsabagh2020firmscope} applied an automated analysis system named \textit{FirmScope} over two thousand firmware images to uncover privilege escalation vulnerabilities in pre-installed apps. Similarly, Zhang et al.~\cite{zhang2022pitracker} implemented an automated tool called \textit{PITracker} to detect insecure intent vulnerabilities in Android pre-installed apps.

% \textbf{Proprietary Vendor Blobs: } Limited research has been conducted to understand the security implications of vendor blobs, primarily because they are distributed as closed-source binaries and lack documentation. This study focuses particularly on closed-source vendor GPU drivers, as they are loaded into the memory space of every app, thus having a significant impact on the security of the Android system.

\section{Conclusion}
This paper presents an empirical study on the security implications of proprietary vendor blobs within Android images. 
In particular, we focus on GPU vendor blobs, as they are loaded into every app's memory space and expose many exploitation code paths.
We design an automatic analysis pipeline and a performant fuzzer to understand their update pattern and uncover  security vulnerabilities. We investigate over 13,000 Android firmware images released between January 2018 and April 2024. Our study reveals that device manufacturers often neglect vendor blob updates during a device's life cycle. Approximately 82\% of firmware images contain outdated GPU blobs (up to 1,128 days). Additionally, a significant number of these vendor blobs are constructed on an obsolete LLVM library released almost 15 years ago. This exposes numerous security vulnerabilities and poses immediate threats to mobile devices. Our work highlights the urgency for timely updates of these vendor blobs by device manufacturers.

\bibliographystyle{unsrt}
\bibliography{cite}

\end{document}